\documentclass[wssquare]{ws-procs9x6} 
\usepackage{physics}
\begin{document}
\begin{flushright}
KEK-TH-2393
\end{flushright}
\vspace{-1cm}

\title{Complex Langevin studies of the emergent space--time\\
in the type IIB matrix model}

\author{Kohta~Hatakeyama$^{1)}$, Konstantinos~Anagnostopoulos$^{2)}$, Takehiro~Azuma$^{3)}$, Mitsuaki~Hirasawa$^{4)}$, Yuta~Ito$^{5)}$, Jun~Nishimura$^{1, 6)}$, Stratos~Papadoudis$^{2)}$, and Asato~Tsuchiya$^{7)}$}

\address{
$^{1)}$ {Theory Center, Institute of Particle and Nuclear Studies,\\
  High Energy Accelerator Research Organization (KEK),\\
  1-1 Oho, Tsukuba, Ibaraki 305-0801, Japan\\
  E-mail: khat@post.kek.jp, jnishi@post.kek.jp}\\
  \vspace{0.15cm}
$^{2)}$ {Physics Department, School of Applied Mathematical and Physical Sciences, National Technical University of Athens, Zografou Campus,\\
  GR-15780 Athens, Greece\\
  E-mail: konstant@mail.ntua.gr, sp10018@central.ntua.gr}\\
  \vspace{0.15cm}
$^{3)}$ {Setsunan University,\\
17-8 Ikeda Nakamachi, Neyagawa, Osaka, 572-8508, Japan\\
  E-mail: azuma@mpg.setsunan.ac.jp}\\
  \vspace{0.15cm}
$^{4)}$ {Sezione di Milano-Bicocca, Istituto Nazionale di Fisica Nucleare (INFN),\\
Piazza della Scienza 3, I-20126 Milano, Italy\\
  E-mail: Mitsuaki.Hirasawa@mib.infn.it}\\
  \vspace{0.15cm}
$^{5)}$ {National Institute of Technology, Tokuyama College,\\
Gakuendai, Shunan, Yamaguchi 745-8585, Japan\\
  E-mail: y-itou@tokuyama.ac.jp}\\
  \vspace{0.15cm}
$^{6)}$ {Department of Particle and Nuclear Physics,\\
School of High Energy Accelerator Science,\\
Graduate University for Advanced Studies (SOKENDAI),\\
1-1 Oho, Tsukuba, Ibaraki 305-0801, Japan}\\
  \vspace{0.15cm}
$^{7)}$ {Department of Physics, Shizuoka University,\\
836 Ohya, Suruga-ku, Shizuoka 422-8529, Japan\\
  E-mail: tsuchiya.asato@shizuoka.ac.jp}
}

\begin{abstract}
The type IIB matrix model has been proposed as a non-perturbative definition of superstring theory since 1996.
We study a simplified model that describes the late time behavior of the type IIB matrix model non-perturbatively using Monte Carlo methods, and we use the complex Langevin method to overcome the sign problem.
We investigate a scenario where the space--time signature changes dynamically from Euclidean at early times to Lorentzian at late times. We discuss the possibility of the emergence of the (3+1)D expanding universe.
\end{abstract}


\bodymatter

\section{Introduction}
\label{sec: intro}
Superstring theory is the most promising candidate for a unified theory of all interactions, including quantum gravity.
The theory is consistently defined in ten-dimensional space-time, leading to the compacting of the extra dimensions into small compact internal spaces.
These scenarios have been investigated perturbatively on D-brane backgrounds and result in a vast number of vacua, leading to the so-called string landscape.
It is, therefore, interesting to see what happens when one includes non-perturbative effects and whether these play an essential role in determining the true vacuum of the theory.
The type IIB matrix model \cite{Ishibashi:1996xs} has been proposed as a non-perturbative definition of superstring theory and provides a promising context to study such problems.

The type IIB matrix model is formally obtained by the dimensional reduction of ten-dimensional ${\cal N}=1$ Super Yang--Mills (SYM)
to zero dimensions.  The theory has maximal ${\cal N}=2$ supersymmetry (SUSY), where translations are realized by the shifts $A_\mu
\to A_\mu + \alpha_\mu \mathbf{1}$, $\mu=0,\ldots,9$.  The eigenvalues of the bosonic matrices $A_\mu$ can therefore be interpreted
as coordinates of space--time.  Thus, in this model, space--time appears dynamically from the degrees of freedom of matrices.  In
the Euclidean version of the model, the Spontaneous Symmetry Breaking (SSB) of the SO(10) rotational symmetry down to SO(3) occurs,
which implies the emergence of a three-dimensional space
\cite{Nishimura:2001sx,Kawai:2002jk,Aoyama:2006rk,Nishimura:2011xy,Anagnostopoulos:2013xga,Anagnostopoulos:2017gos,Anagnostopoulos:2020xai}.

By Monte Carlo simulation \cite{Kim:2011cr}, it was found that a continuous time emerges dynamically, and a three-dimensional space expands.
In Refs.~\cite{Ito:2013ywa,Ito:2015mxa}, it turned out that the expanding behavior of the space obeys the exponential law at early times and the power-law at late times. 
In Ref.~\cite{Aoki:2019tby}, however, it was shown that SSB comes from singular configurations associated with the Pauli matrices, in which only two eigenvalues are large.
This problem has been attributed to an approximation used to avoid the sign problem, which turned out later to be unjustifiable.

In Refs.~\cite{Nishimura:2019qal,Hatakeyama:2021ake,Hirasawa:2021xeh}, the Complex Langevin Method (CLM) \cite{Parisi:1983mgm,Klauder:1983sp} was used to overcome the sign problem without the above mentioned approximation.
When one applies this method, one should apply the criterion for correct convergence of the CLM \cite{Aarts:2009dg,Aarts:2009uq,Aarts:2011ax,Nishimura:2015pba,Nagata:2015uga,Nagata:2016vkn,Ito:2016efb}.
In Ref.~\cite{Hirasawa:2021xeh}, we found a new phase in which the structure of space is continuous by applying the CLM to the Lorentzian type IIB matrix model.
See also Refs.~\cite{Brahma:2021tkh,Steinacker:2021yxt,Klinkhamer:2021nyt} for other related works.

In this work, we study the bosonic version of the type IIB matrix model by using the CLM.
We show the equivalence between the Lorentzian and Euclidean models, which implies that the space--time in the Lorentzian model is Euclidean.
To realize the possibility of the dynamical change of signature from Euclidean to Lorentzian, we introduce a Lorentz-invariant mass term in the action that breaks the equivalence.
We find some evidence that the signature of space--time changes from Euclidean at early times to Lorentzian at later times.

\section{The type IIB matrix model}
\label{sec: IIBMM}

\subsection{Definition}
\label{sec: def_IIBMM}
The action of the type IIB matrix model is given as follows: $S= S_\mathrm{b}+S_\mathrm{f}$,
\begin{equation}
\label{eq: action}
S_\mathrm{b}= -\frac{1}{4g^2} \Tr \qty( [A^\mu,A^\nu][A_\mu,A_\nu]) \ ,\quad
S_\mathrm{f}= -\frac{1}{2g^2} \Tr \qty(\bar{\Psi}(\mathcal{C}\Gamma^\mu) [A_\mu,\Psi])\ ,
\end{equation}
where $A_\mu \ (\mu=0,\ldots,9)$ and $\Psi$ are $N\times N$ Hermitian matrices, and $\Gamma^\mu$ and $\mathcal{C}$ are
10-dimensional gamma matrices and the charge conjugation matrix, respectively, which are obtained after the Weyl projection.
The $A_\mu$ and $\Psi$ transform as vectors and Majorana-Weyl spinors under SO(9,1) transformations.
In this study, we omit $S_\mathrm{f}$ to reduce the computational cost.

The partition function is given by $Z=\int dA e^{iS_\mathrm{b}}$.
Due to the phase factor $e^{iS_\mathrm{b}}$, the model is not well-defined as it is, and in this work, we define it by deforming the integration contour.
When we rewrite the partition function as $Z=\int dA e^{-\tilde{S}}$, the action of the Lorentzian model is given as
\begin{equation}
\tilde{S}=-\frac{i}{4}N\qty[-2\Tr({F}_{0i})^2 +\Tr({F}_{ij})^2]\ ,
\end{equation}
where $g^2=1/N$ and $F_{\mu\nu}=i\comm{A_\mu}{A_\nu}$.
According to Cauchy's theorem, one can rotate the Lorentzian matrices $A_\mu$ to the Euclidean ones $\tilde{A}_\mu$ since the integration contour of $A_\mu$ can be deformed keeping the real part of $\tilde{S}$ positive.
The relationship between $A_\mu$ and $\tilde{A}_\mu$ is
\begin{equation}
\label{eq: rotation_A}
A_0=e^{-i\frac{3\pi}{8}}\tilde{A}_0 \ , \quad A_i=e^{i\frac{\pi}{8}}\tilde{A}_i\ .
\end{equation}
Then, the Euclidean action is given by
\begin{equation}
\tilde{S} = \frac{1}{4} N \qty[2 \text{Tr}(\tilde F_{0i})^2 +\text{Tr}(\tilde F_{ij})^2]\ ,
\end{equation}
which is positive-definite.
Here we have defined $\tilde F_{\mu\nu}=i [\tilde{A}_\mu, \tilde{A}_\nu]$.

\subsection{Equivalence between the Euclidean and Lorentzian models}

\begin{figure}
\centering
\includegraphics[scale=0.37]{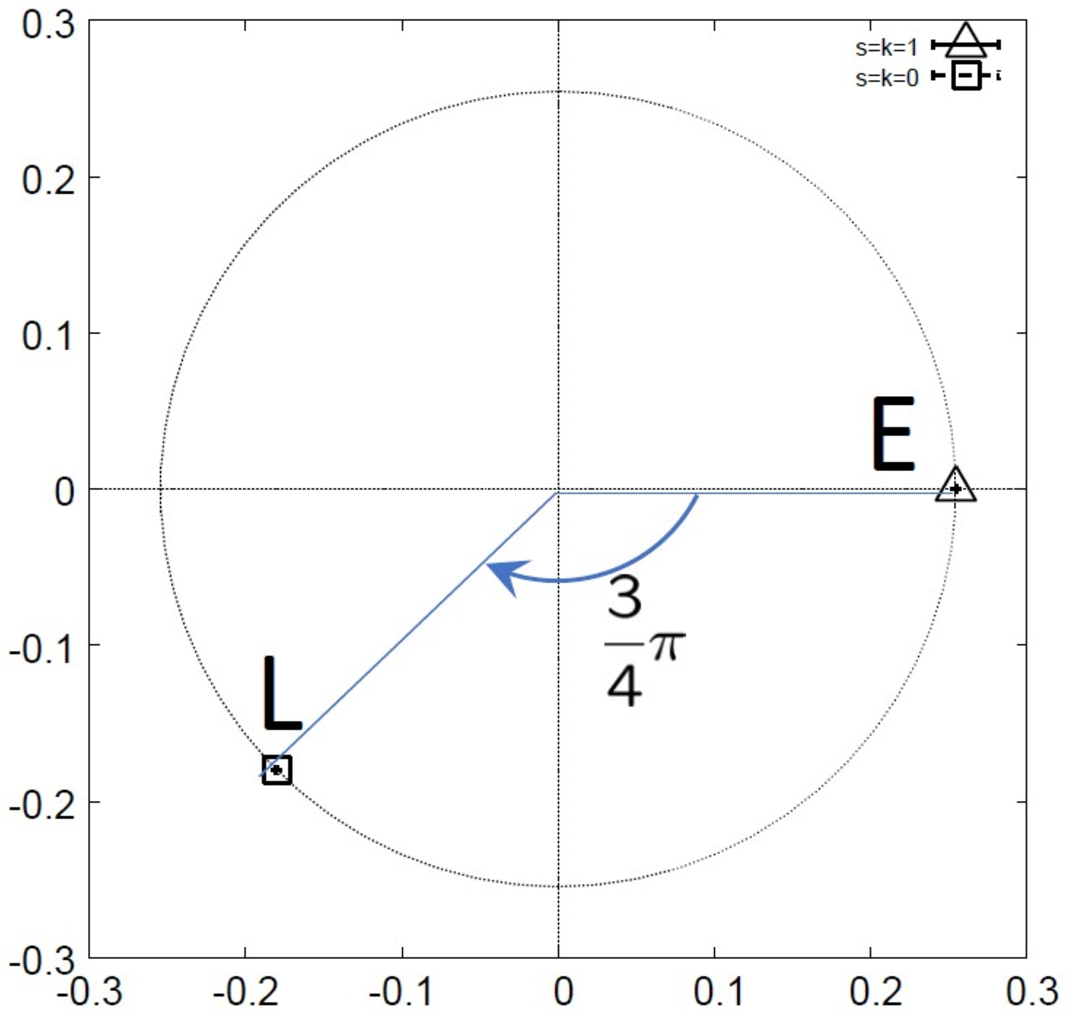}
\includegraphics[scale=0.37]{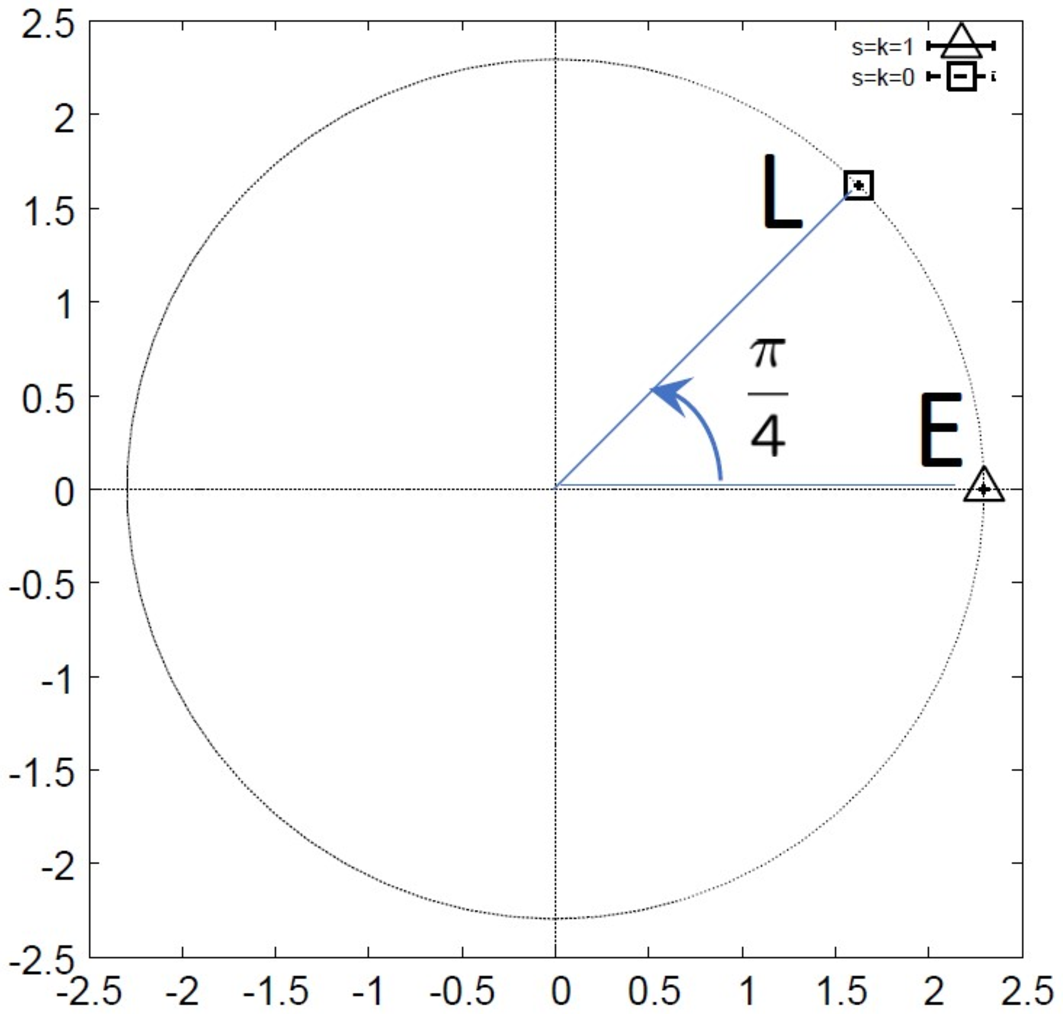}
\caption{We plot the expectation values of $\frac{1}{N}\Tr A_0^2$ (Left) and $\frac{1}{N}\Tr A_i^2$ (Right).
Those of the Euclidean and Lorentzian models are represented by the triangles and the squares, respectively.
The angles between the Lorentzian and Euclidean models of $\langle \frac{1}{N} \text{Tr} \tilde{A}_0^2 \rangle$ and $\langle \frac{1}{N} \text{Tr} \tilde{A}_i^2 \rangle$ are $-3\pi/4$ and $\pi/4$, which agree with Eq.~\eqref{eq: TrAsq}.
}
\label{fig: TrAisq}
\end{figure}
By using Eq.~\eqref{eq: rotation_A}, one can derive the relationship between the expectation values of $\Tr A_0^2$ and $\Tr A_i^2$ in the two models:
\begin{equation}
\label{eq: TrAsq}
\expval{\frac{1}{N} \text{Tr} A_0^2}_\mathrm{L}
=e^{-i\frac{3\pi}{4}}\expval{\frac{1}{N} \text{Tr} \tilde{A}_0^2}_\mathrm{E} \ ,
\quad
\expval{\frac{1}{N} \text{Tr} A_i^2}_\mathrm{L}
=e^{i\frac{\pi}{4}}\expval{\frac{1}{N} \text{Tr} \tilde{A}_i^2}_\mathrm{E}\ ,
\end{equation}
where $\expval{\ \cdot \ }_\mathrm{L}$ and $\expval{\ \cdot \ }_\mathrm{E}$ denote the expectation values in the Lorentzian and Euclidean models, respectively.
In Fig.~\ref{fig: TrAisq} (Left), $\langle \frac{1}{N} \text{Tr} A_0^2 \rangle$ is shown, and the angle between $\langle \frac{1}{N} \text{Tr} \tilde{A}_0^2 \rangle_\mathrm{E}$ and $\langle \frac{1}{N} \text{Tr} A_0^2 \rangle_\mathrm{L}$ is $-3\pi/4$.
In Fig.~\ref{fig: TrAisq} (Right), $\langle \frac{1}{N} \text{Tr} A_i^2 \rangle$ is shown, and the angle between  $\langle \frac{1}{N} \text{Tr} \tilde{A}_i^2 \rangle_\mathrm{E}$ and $\langle \frac{1}{N} \text{Tr} A_i^2 \rangle_\mathrm{L}$ is $\pi/4$.
These angles are in agreement with Eq.~\eqref{eq: TrAsq}.

These results are consistent with the fact that the Lorentzian and the Euclidean models are equivalent. Expectation values in the Lorentzian model can be obtained by simply rotating the phase of those in the Euclidean model.
In particular, $\langle \frac{1}{N} \text{Tr} A_0^2 \rangle_\mathrm{L}$ and $\langle \frac{1}{N} \text{Tr} A_i^2 \rangle_\mathrm{L}$ are complex and the emergent space--time should be interpreted as Euclidean.

\subsection{Lorentz-invariant mass term}
To realize real time and space, we introduce a {\it Lorentz-invariant mass term} in the action.
For the Lorentzian model, the action is
\begin{equation}
\label{eq: mass_term_Lo}
\tilde{S}=-\frac{i}{4}N\qty[-2\Tr({F}_{0i})^2 +\Tr({F}_{ij})^2] -\frac{i}{2}N\gamma\qty[\Tr(A_0)^2 -\Tr(A_i)^2]
\end{equation}
with $\gamma > 0$.
Using Eq.~\eqref{eq: rotation_A}, we find that the action for the corresponding Euclidean model becomes
\begin{equation}
\label{eq: mass_term_Eu}
\tilde{S} = \frac{1}{4}N \qty[2\text{Tr}(\tilde F_{0i})^2 +\text{Tr}(\tilde F_{ij})^2]
+\frac{1}{2}N\gamma\; e^{i\frac{3\pi}{4}}\qty[\text{Tr}(\tilde{A}_0)^2 +\text{Tr}(\tilde{A}_i)^2]\ ,
\end{equation}
where the real part of the mass term is negative.
If $\gamma < 0$, the real part of the mass term in the Euclidean model becomes positive, and then the matrices can be rotated from the Lorentzian to the Euclidean, which implies the equivalence between the two models.

The same mass term was used to study classical solutions of the Lorentzian type IIB matrix model \cite{Hatakeyama:2019jyw}:
\begin{equation}
[A^\nu, [A_\nu, A_\mu]]-\gamma A_\mu=0 \ .
\end{equation}
For $\gamma>0$, one can obtain classical solutions with smooth space and expanding behavior.
Classical solutions with Hermitian $A_\mu$ make the time and space real.
For $\gamma=0$, the classical solutions are given by simultaneously diagonalizable $A_\mu$, which do not necessarily have an expanding behavior.
For $\gamma<0$, there do not exist classical solutions with expanding behavior.
\subsection{The time evolution}
\label{sec: time_evolution}
As mentioned in Sec.~\ref{sec: intro}, time does not exist a priori, and we define it as follows.
We choose a basis in which $A_0$ is diagonal and its eigenvalues are in the ascending order: $A_0=\text{diag}(\alpha_1,\alpha_2,\ldots,\alpha_N)\, ,\quad \alpha_1\le\alpha_2\le\dots\le\alpha_N$.
Then, we define $\bar{\alpha}_k$ as $\bar{\alpha}_k = \frac{1}{n} \sum_{i=1}^{n} \alpha_{k+i}$, and the time $t_\rho$ as
\begin{equation}
\label{eq: time}
t_\rho = \sum_{k=1}^{\rho} \abs{\bar{\alpha}_{k+1} -  \bar{\alpha}_k } \ .
\end{equation}
Here, we introduce the $n \times n$ matrices $\bar{A}_i(t)$ as $\qty(\bar{A}_i)_{ab}(t) = \qty(A_i)_{k+a, k+b}$, which represent the space at the time $t$.

\section{Complex Langevin method}
The complex Langevin method (CLM) \cite{Parisi:1983mgm, Klauder:1983sp} can be applied successfully to many systems with a complex action problem.
One writes down stochastic differential equations for the complexified degrees of freedom, which can be used to compute expectation values under certain conditions.
Consider a model given by the partition function $Z=\int dx\, w(x)$, where $x\in \mathbb{R}^n$ and $w(x)$ is a complex-valued function.
In the CLM, we complexify the variables $x\in \mathbb{R}^n \longrightarrow z \in \mathbb{C}^n$, and solve the complex Langevin equation with the Langevin time $\sigma$:
\begin{equation}
\label{Langevin_eq}
\frac{dz_k}{d\sigma}={\frac{1}{w(z)}\pdv{w(z)}{z_k}} +{\eta_k(\sigma)}\ .
\end{equation}
The first term of the right-hand side of Eq.~\eqref{Langevin_eq} is the drift term, and the second one is the real Gaussian noise with the probability distribution
\begin{equation}
\mathrm{P}(\eta_k(\sigma)) \propto e^{-\frac{1}{4}\int d\sigma \sum_k [\eta_k(\sigma)]^2}\ .
\end{equation}
To confirm that the CLM gives correct solutions, we use the criterion that the probability distribution of the drift term should be exponentially suppressed for large values \cite{Nagata:2016vkn}.

\subsection{Application of the CLM to the type IIB matrix model}
To apply the CLM to the type IIB matrix model, we make a change of variables \cite{Nishimura:2019qal}: $\alpha_1=0\ ,\ \alpha_i = \sum_{k=1}^{i-1} e^{\tau_k}$ for $2 \leq i \leq N$, where we introduce new real variables $\tau_k$. 
In this way, the ordering of $\alpha_i$ is automatically realized.
Initially, $\alpha_i$ are real, and $A_i$ are Hermitian matrices. To apply the CLM, we complexify  $\tau_k$ and take $A_i$ to be SL($N,\mathbb{C}$) matrices.
The complex Langevin equations are given by
\begin{equation}
\frac{d\tau_k}{d\sigma}
=-{\pdv{S_\mathrm{eff}}{\tau_k}} +{\eta_k(\sigma)} \ ,
\quad
\frac{d(A_i)_{kl}}{d\sigma}
=-{\pdv{S_\mathrm{eff}}{(A_i)_{lk}}} +{(\eta_i)_{kl}(\sigma)}\ ,
\end{equation}
where $S_\mathrm{eff}$ is obtained from $\tilde{S}$ in Eq.~\eqref{eq: mass_term_Lo} by adding a term associated with the gauge fixing and the Jacobian term associated with the change of variables.

\section{Results}
In the following, we introduce a parameter $\varepsilon$ in the mass term:
\begin{equation}
\tilde{S} = -\frac{i}{4}N\qty[-2\Tr(F_{0i})^2 +\Tr(F_{ij})^2]
-\frac{i}{2}N\gamma\qty[{e^{i\varepsilon}}\Tr(A_0)^2 -{e^{-i\varepsilon}}\Tr(A_i)^2]
\end{equation}
to shift coefficients of $\Tr(A_0)^2$ and $\Tr(A_i)^2$ slightly from pure imaginary, and set $\varepsilon=\pi/10$. 

\subsection{Expectation value of the time coordinate}
\begin{figure}
\centering
\includegraphics[scale=0.46]{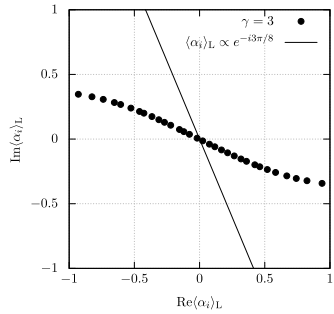}
\caption{Expectation values of the eigenvalues $\alpha_i$ of $A_0$ for $N=32,\gamma=3$ are plotted.
The solid line corresponds to the Euclidean model, where the complex phase of the expectation values $\expval{\alpha_i}_\mathrm{L}$ is $\exp(-i3\pi/8)$.
From this plot, $\theta_\mathrm{t}$ tends to become 0 at late times (at both ends of the distribution).
}
\label{fig: alpha}
\end{figure}

When $\gamma=0$, Eq.~\eqref{eq: rotation_A} holds, and we expect that $\expval{\alpha_i}_\mathrm{L} =e^{-i\frac{3\pi}{8}}\expval{\tilde{\alpha}_i}_\mathrm{E}$.
This is true because of the equivalence between the Euclidean and Lorentzian models, and time is regarded as the Euclidean one.
We measure the time differences $(\Delta \alpha_i)_\mathrm{L}= (\alpha_{i+1})_\mathrm{L} -(\alpha_i)_\mathrm{L} \propto e^{i\theta_\mathrm{t}}$.
The emergent time is real if $\theta_\mathrm{t}=0$.

In Fig.~\ref{fig: alpha}, we plot the expectation values of the time coordinates $\expval{\alpha_i}_\mathrm{L}$ on the complex plane for $N=32, \gamma=3$.
The solid line corresponds to the Euclidean model, where the complex phase of $\expval{\alpha_i}_\mathrm{L}$ is $\exp(-i3\pi/8)$.
From the plot, $\theta_\mathrm{t}$ tends to become $0$ at late times (at both ends of the distribution).

\subsection{Time evolution of space}
%
\begin{figure}
\centering
\includegraphics[scale=0.37]{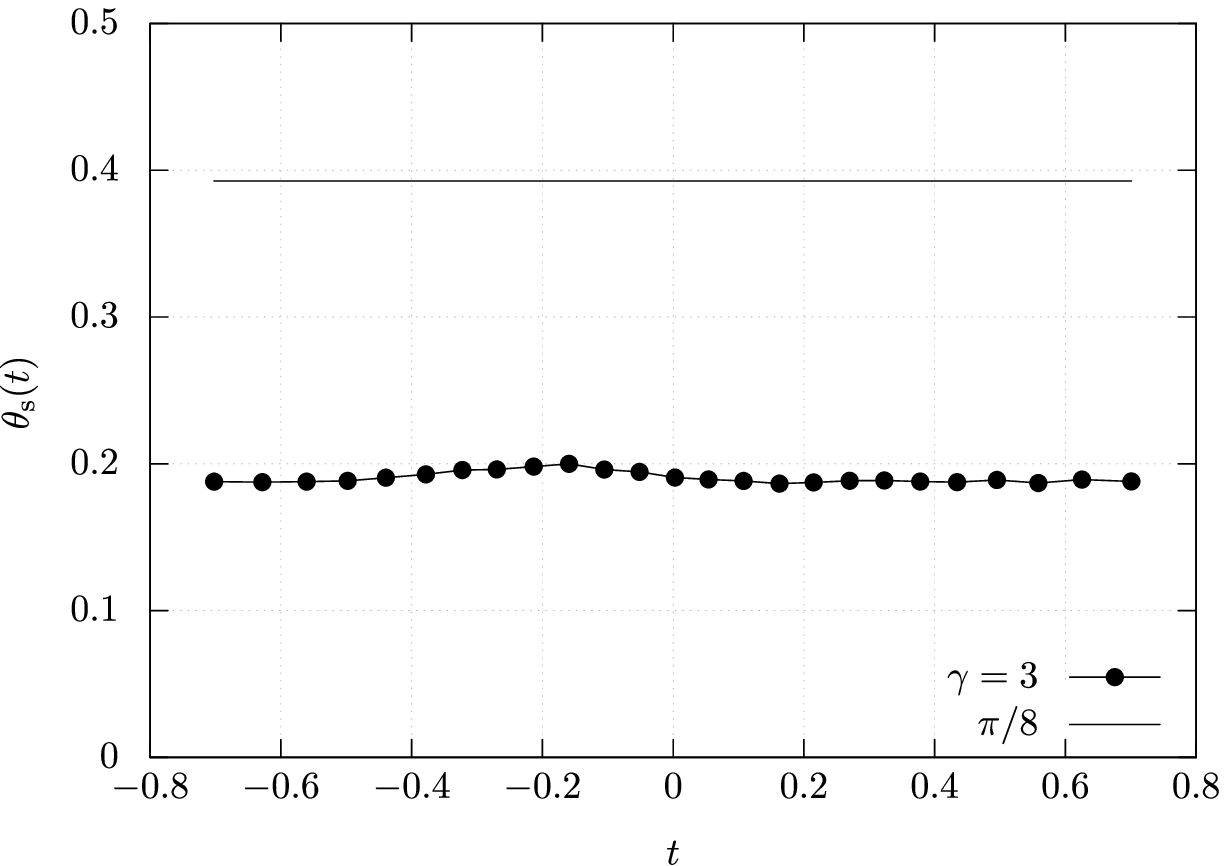}
\includegraphics[scale=0.37]{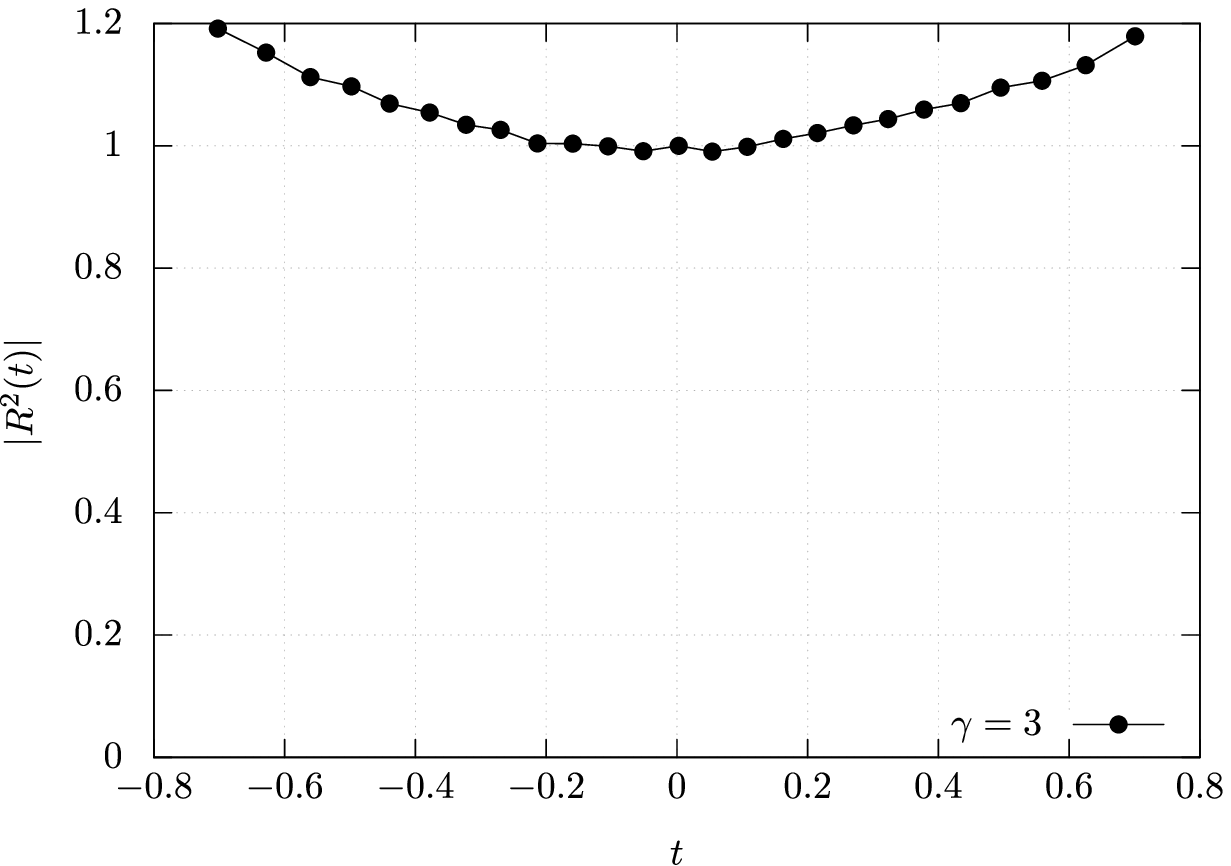}
\caption{(Left) $\theta_\mathrm{s}(t)$ is plotted against $t$ for $\gamma=3$.
All values of $\theta_\mathrm{s}(t)$ are about $0.2$ and below the $\theta_\mathrm{s}(t)=\pi/8$ line, which corresponds to the Euclidean space.
(Right) $|R^2(t)|$ is plotted against $t$ for $\gamma=3$.
We can see that the space is expanding slightly with the time $t$.}

\label{fig: Rsq}
\end{figure}
The time evolution of the extent of space is given by $R^2(t) = \expval{\frac{1}{n} \tr \qty(\bar{A}_i(t))^2} =e^{2i\theta_\mathrm{s}(t)}\abs{R^2(t)}$.
Since the matrices  $\bar{A}_i$ are complex, $R^2(t)$ is also complex.
The time $t$ is defined in Eq.~\eqref{eq: time}.
From Eq.~\eqref{eq: TrAsq}, we obtain the Euclidean space when $\theta_\mathrm{s}(t) \sim \pi/8$, and the real space in the Lorentzian model when $\theta_\mathrm{s}(t) \sim 0$.
Therefore, the signature of space--time can change dynamically in this model.

In Fig.~\ref{fig: Rsq} (Left) and (Right), $\theta_\mathrm{s}(t)$ and $|R^2(t)|$ are plotted against $t$ for $N=32, \gamma=3$, respectively.
All values of $\theta_\mathrm{s}(t)$ are about $0.2$ and below the $\theta_\mathrm{s}(t)=\pi/8$ line, which corresponds to the Euclidean space.
We can see that the space is expanding slightly with the time $t$ from the plot of $|R^2(t)|$.

\section{Conclusions}
In this work, the CLM was applied to the bosonic type IIB matrix model in order to overcome the sign problem.
We showed that the Lorentzian and Euclidean models are equivalent and that expectation values in the two models are related to each other by some complex phase rotation.
The expectation values \eqref{eq: TrAsq} in the Lorentzian model are complex, and space--time is Euclidean.

We introduced the model with a Lorentz-invariant mass term, which is a promising way to realize real time and expanding space.
Then, the Euclidean and Lorentzian models are not equivalent anymore for $\gamma>0$.
We found that the time, which is extracted from the expectation values of the eigenvalues of $A_0$ in the Lorentzian model, may be real at late times although they are complex near the origin.
We also studied the evolution of the extent of space with time.
We have seen some tendency that the space becomes closer to real than the original model.

To obtain a three-dimensional expanding space, we expect that supersymmetry will play an essential role. 
We are currently investigating its effect, which we will report in the near future.

\section*{Acknowledgments}
T.\;A., K.\;H. and A.\;T. were supported in part by Grant-in-Aid (Nos. 17K05425, 19J10002, and 18K03614, 21K03532, respectively) from Japan Society for the Promotion of Science.
This research was supported by MEXT as ``Program for Promoting Researches on the Supercomputer Fugaku'' (Simulation for basic science: from fundamental laws of particles to creation of nuclei, JPMXP1020200105) and JICFuS.
This work used computational resources of supercomputer Fugaku provided by the RIKEN Center for Computational Science (Project ID: hp210165) and Oakbridge-CX provided by the University of Tokyo (Project IDs: hp200106, hp200130, hp210094) through the HPCI System Research Project.
Numerical computation was also carried out on PC clusters in KEK Computing Research Center.
This work was also supported by computational time granted by the Greek Research and Technology Network (GRNET) in the National HPC facility ARIS, under the project IDs SUSYMM and SUSYMM2.

\bibliographystyle{ws-procs9x6} 

\bibliography{ref_CLMIKKT}

\begin{thebibliography}{10}

\bibitem{Ishibashi:1996xs}
N.~Ishibashi, H.~Kawai, Y.~Kitazawa and A.~Tsuchiya, {A Large N reduced model
  as superstring}, {\em Nucl. Phys. B} {\bf 498}, 467  (1997).

\bibitem{Nishimura:2001sx}
J.~Nishimura and F.~Sugino, {Dynamical generation of four-dimensional
  space-time in the IIB matrix model}, {\em JHEP} {\bf 05}, 001  (2002).

\bibitem{Kawai:2002jk}
H.~Kawai, S.~Kawamoto, T.~Kuroki, T.~Matsuo and S.~Shinohara, {Mean field
  approximation of IIB matrix model and emergence of four-dimensional
  space-time}, {\em Nucl. Phys. B} {\bf 647}, 153  (2002).

\bibitem{Aoyama:2006rk}
T.~Aoyama and H.~Kawai, {Higher order terms of improved mean field
  approximation for IIB matrix model and emergence of four-dimensional
  space-time}, {\em Prog. Theor. Phys.} {\bf 116}, 405  (2006).

\bibitem{Nishimura:2011xy}
J.~Nishimura, T.~Okubo and F.~Sugino, {Systematic study of the SO(10) symmetry
  breaking vacua in the matrix model for type IIB superstrings}, {\em JHEP}
  {\bf 10}, 135  (2011).

\bibitem{Anagnostopoulos:2013xga}
K.~N. Anagnostopoulos, T.~Azuma and J.~Nishimura, {Monte Carlo studies of the
  spontaneous rotational symmetry breaking in dimensionally reduced super
  Yang-Mills models}, {\em JHEP} {\bf 11}, 009  (2013).

\bibitem{Anagnostopoulos:2017gos}
K.~N. Anagnostopoulos, T.~Azuma, Y.~Ito, J.~Nishimura and S.~K. Papadoudis,
  {Complex Langevin analysis of the spontaneous symmetry breaking in
  dimensionally reduced super Yang-Mills models}, {\em JHEP} {\bf 02}, 151
  (2018).

\bibitem{Anagnostopoulos:2020xai}
K.~N. Anagnostopoulos, T.~Azuma, Y.~Ito, J.~Nishimura, T.~Okubo and
  S.~Kovalkov~Papadoudis, {Complex Langevin analysis of the spontaneous
  breaking of 10D rotational symmetry in the Euclidean IKKT matrix model}, {\em
  JHEP} {\bf 06}, 069  (2020).

\bibitem{Kim:2011cr}
S.-W. Kim, J.~Nishimura and A.~Tsuchiya, {Expanding (3+1)-dimensional universe
  from a Lorentzian matrix model for superstring theory in (9+1)-dimensions},
  {\em Phys. Rev. Lett.} {\bf 108}, 011601  (2012).

\bibitem{Ito:2013ywa}
Y.~Ito, S.-W. Kim, Y.~Koizuka, J.~Nishimura and A.~Tsuchiya, {A renormalization
  group method for studying the early universe in the Lorentzian IIB matrix
  model}, {\em PTEP} {\bf 2014}, 083B01  (2014).

\bibitem{Ito:2015mxa}
Y.~Ito, J.~Nishimura and A.~Tsuchiya, {Power-law expansion of the Universe from
  the bosonic Lorentzian type IIB matrix model}, {\em JHEP} {\bf 11}, 070
  (2015).

\bibitem{Aoki:2019tby}
T.~Aoki, M.~Hirasawa, Y.~Ito, J.~Nishimura and A.~Tsuchiya, {On the structure
  of the emergent 3d expanding space in the Lorentzian type IIB matrix model},
  {\em PTEP} {\bf 2019}, 093B03  (2019).

\bibitem{Nishimura:2019qal}
J.~Nishimura and A.~Tsuchiya, {Complex Langevin analysis of the space-time
  structure in the Lorentzian type IIB matrix model}, {\em JHEP} {\bf 06}, p.
  077  (2019).

\bibitem{Hatakeyama:2021ake}
K.~Hatakeyama, K.~Anagnostopoulos, T.~Azuma, M.~Hirasawa, Y.~Ito, J.~Nishimura,
  S.~Papadoudis and A.~Tsuchiya, {Relationship between the Euclidean and
  Lorentzian versions of the type IIB matrix model}, in {\em {38th
  International Symposium on Lattice Field Theory}\/},  {\em
  \verb|arXiv:2112.15368 [hep-lat]|}.

\bibitem{Hirasawa:2021xeh}
M.~Hirasawa, K.~Anagnostopoulos, T.~Azuma, K.~Hatakeyama, Y.~Ito, J.~Nishimura,
  S.~Papadoudis and A.~Tsuchiya, {A new phase in the Lorentzian type IIB matrix
  model and the emergence of continuous space-time}, in {\em {38th
  International Symposium on Lattice Field Theory}\/},  {\em
  \verb|arXiv:2112.15390 [hep-lat]|}.

\bibitem{Parisi:1983mgm}
G.~Parisi, {ON COMPLEX PROBABILITIES}, {\em Phys. Lett. B} {\bf 131}, 393
  (1983).

\bibitem{Klauder:1983sp}
J.~R. Klauder, {Coherent State Langevin Equations for Canonical Quantum Systems
  With Applications to the Quantized Hall Effect}, {\em Phys. Rev. A} {\bf 29},
  2036  (1984).

\bibitem{Aarts:2009dg}
G.~Aarts, F.~A. James, E.~Seiler and I.-O. Stamatescu, {Adaptive stepsize and
  instabilities in complex Langevin dynamics}, {\em Phys. Lett. B} {\bf 687},
  154  (2010).

\bibitem{Aarts:2009uq}
G.~Aarts, E.~Seiler and I.-O. Stamatescu, {The Complex Langevin method: When
  can it be trusted?}, {\em Phys. Rev. D} {\bf 81}, 054508  (2010).

\bibitem{Aarts:2011ax}
G.~Aarts, F.~A. James, E.~Seiler and I.-O. Stamatescu, {Complex Langevin:
  Etiology and Diagnostics of its Main Problem}, {\em Eur. Phys. J. C} {\bf
  71}, 1756  (2011).

\bibitem{Nishimura:2015pba}
J.~Nishimura and S.~Shimasaki, {New Insights into the Problem with a Singular
  Drift Term in the Complex Langevin Method}, {\em Phys. Rev. D} {\bf 92}, 011501  (2015).

\bibitem{Nagata:2015uga}
K.~Nagata, J.~Nishimura and S.~Shimasaki, {Justification of the complex
  Langevin method with the gauge cooling procedure}, {\em PTEP} {\bf 2016}, 013B01  (2016).

\bibitem{Nagata:2016vkn}
K.~Nagata, J.~Nishimura and S.~Shimasaki, {Argument for justification of the
  complex Langevin method and the condition for correct convergence}, {\em
  Phys. Rev. D} {\bf 94}, 114515  (2016).

\bibitem{Ito:2016efb}
Y.~Ito and J.~Nishimura, {The complex Langevin analysis of spontaneous symmetry
  breaking induced by complex fermion determinant}, {\em JHEP} {\bf 12}, 009
   (2016).

\bibitem{Brahma:2021tkh}
S.~Brahma, R.~Brandenberger and S.~Laliberte, {Emergent Cosmology from Matrix
  Theory}, {\em \verb|arXiv:2107.11512 [hep-th]|}.

\bibitem{Steinacker:2021yxt}
H.~C. Steinacker, {Gravity as a Quantum Effect on Quantum Space-Time}, {\em
  \verb|arXiv:2110.03936 [hep-th]|}.

\bibitem{Klinkhamer:2021nyt}
F.~R. Klinkhamer, {Towards a numerical solution of the bosonic master-field
  equation of the IIB matrix model}, {\em \verb|arXiv:2110.15309 [hep-th]|}.

\bibitem{Hatakeyama:2019jyw}
K.~Hatakeyama, A.~Matsumoto, J.~Nishimura, A.~Tsuchiya and A.~Yosprakob, {The
  emergence of expanding space\textendash{}time and intersecting D-branes from
  classical solutions in the Lorentzian type IIB matrix model}, {\em PTEP} {\bf
  2020}, 043B10  (2020).

\end{thebibliography}

\end{document}